\documentclass[aps,prd,preprintnumbers,nofootinbib]{revtex4}
\usepackage{amsmath}
\usepackage{amssymb}
\usepackage{latexsym}
\usepackage{graphics}
\usepackage{psfrag,epsfig}
\usepackage{hyperref}
\def\beq{\begin{equation}}
\def\eeq{\end{equation}}
\def\bea{\begin{eqnarray}}
\def\eea{\end{eqnarray}}
\def\bean{\begin{eqnarray*}}
\def\eean{\end{eqnarray*}}
\begin{document}

\title{Diagrammatic Derivation of Lovelock Densities}

\author{\href{mailto:Charalampos.Bogdanos@th.u-psud.fr}{C. Bogdanos}}

\affiliation{ \href{http://www.th.u-psud.fr/}{Laboratoire de Physique Theorique} \\
 \href{http://www.u-psud.fr}{Universit\'e de Paris-Sud 11} \\
 B\^atiment 210, 91405 Orsay CEDEX, France}

\begin{abstract}
We discuss a method of calculating the various scalar densities encountered in Lovelock theory which relies on diagrammatic, instead of algebraic manipulations. Taking advantage of the known symmetric and antisymmetric properties of the Riemann tensor which appears in the Lovelock densities, we map every quadratic or higher contraction into a corresponding permutation diagram. The derivation of the explicit form of each density is then reduced to identifying the distinct diagrams, from which we can also read off the overall combinatoric factors. The method is applied to the first Lovelock densities, of order two (Gauss-Bonnet term) and three. 
\end{abstract}
\maketitle

Lovelock theory is a natural classical extension of General Relativity in higher dimensions~\cite{Lovelock:1971yv}. With the advent of a renewed interest on extra-dimensional models of gravity, particularly through String/M-Theory~\cite{Polchinski:1995mt} and braneworld models~\cite{Akama:1982jy,Rubakov:1983bb,Antoniadis:1998ig,ArkaniHamed:1998rs,Randall:1999ee,Randall:1999vf}, Lovelock gravity has received a lot of attention. The theory assumes the inclusion of an infinite number of scalar densities in the gravitational Lagrangian, apart from the ordinary Einstein-Hilbert term. Although these do not introduce any new dynamics when restricted in four dimensions, they become important when extra dimensions are introduced~\cite{Charmousis:2008kc,Dadhich:2009mp}. The first of these Lovelock densities is the well known Gauss-Bonnet scalar, which also appears in perturbative expansions in String theory, giving further support for the relevance of Lovelock theory~\cite{Zwiebach:1985uq,Zumino:1985dp}. 

However, calculations in Lovelock gravity are known to be quite involved. The inclusion of higher scalar densities leads to considerably more complicated equations and these technical challenges render very difficult a systematic treatment of the theory to all orders. The derivation of even the scalar densities themselves quickly becomes very complicated once we start increasing the order~\cite{MuellerHoissen:1985mm,Briggs:1997ae,Briggs:1996bj}. Higher terms include contractions of  products of two or more Riemann tensors and the possible combinations lead to a combinatoric explosion in complexity. As a result, up to now only the lowest order Lovelock correction, i.e. the Gauss-Bonnet term, has been extensively studied~\cite{Cai:2001dz,Cai:2003gr}, with recent attempts to also include a third and fourth order density~\cite{Rizzo:2006uz,Dehghani:2006yd,HabibMazharimousavi:2008ib,Mazharimousavi:2008ti}.  
 
In this letter we discuss an alternative method of calculating the first Lovelock densities, based on the use of permutation diagrams, instead of the more conventional way of using the Kronecker permutation tensor. We demonstrate that this method can lead to dramatic simplification of certain calculations, rendering the derivation of the first few scalar densities almost trivial. The method relies on the symmetries of the Riemann tensor in order to simplify and group together diagrams that lead to the same expressions. Essentially, the calculation of a scalar density is thus reduced to writing down the corresponding diagrams using a number of rules and then determining from them the overall numerical factor and sign of each term by inspecting the form of the appropriate diagram.We illustrate this procedure for the scalar densities ${\mathcal L}_{(2)}$ (Gauss-Bonnet term) and ${\mathcal L}_{(3)}$.

The Lovelock action in $D$ dimensions can be written as 
\beq
S_D  = \int_{\cal M} {\sum\limits_{k = 0}^{\left\lfloor {\left( {D - 1} \right)/2} \right\rfloor } {\alpha _k {\cal L}_{\left( k \right)} } } \,,
\eeq
with ${\mathcal M}$ being the D-dimensional spacetime manifold, $\alpha_{k}$ are coupling constants and the various Lovelock densities are given from the general expression
\beq
{\cal L}_{\left( k \right)}  = {\cal R}^{A_1 B_1 }  \wedge  \ldots  \wedge {\cal R}^{A_k B_k }  \wedge \theta ^ *  _{A_1 B_1  \ldots A_k B_k } \,,
\eeq
where ${\cal R}^A _{\;B}  = d\omega ^A _{\;B}  + \omega ^A _{\;C}  \wedge \omega ^C _{\;B} $ are the curvature 2-forms. The Hodge dual of a k-form is defined as 
\beq
\theta ^ *  _{A_1  \ldots A_k }  = \frac{1}{{\left( {D - k} \right)!}}\varepsilon _{A_1  \ldots A_k A_{k + 1}  \ldots A_D } \theta ^{A_{k + 1} }  \wedge  \ldots  \wedge \theta ^{A_D } \,.
\eeq
Using the above definitions, one can take contractions and write down each scalar tensity in component notation. The expressions we end up are of the general form~\cite{Deruelle:2003ck}
\beq
R^{A_1 B_1 } _{\quad \quad C_1 D_1 } R^{A_2 B_2 } _{\quad \quad C_2 D_2 }  \ldots R^{A_k B_k } _{\quad \quad C_k D_k } \delta ^{C_1 D_1  \ldots C_k D_k } _{A_1 B_1  \ldots A_k B_k } 
\eeq
where $\delta ^{C_1 D_1  \ldots C_k D_k } _{A_1 B_1  \ldots A_k B_k } $ is the antisymmetric Kronecker symbol, which has the property to be equal to $\pm 1$ whenever $A_{1}B_{1} \dots A_{k}B_{k}$ is an even or odd permutation of $C_{1}D_{1} \dots C_{k}D_{k}$ respectively and zero in any other case. Thus, in order to write a specific density, one has to break down the antisymmetric Kronecker symbol into products of ordinary Kronecker deltas. Obviously, there will be as many terms as there are possible permutations of the indices $A_{1}B_{1} \dots A_{k}B_{k}$ and $C_{1}D_{1} \dots C_{k}D_{k}$. And this is where the calculational problem arises, as the number of available terms quickly escalates with $k$. For ${\cal L}_{(1)}$, which is the ordinary Einstein-Hilbert Lagrangian, we have $2!$ terms, for ${\cal L}_{(2)}$ this increases to $4!=24$ and for ${\cal L}_{(3)}$ we already reach $6!=720$ terms. Clearly, it is not feasible for someone to write each one of 720 terms just to get to $k=3$. This algebraic approach, although quite general in nature, is in the end unsuitable for practical calculations. We can acknowledge the existence of an easier path to the solution, given that the end results are usually considerably simpler compared to the expressions arising in the conventional algebraic derivation. For example, the Gauss-Bonnet scalar only involves 3 distinct terms, which means there is a high amount of redundancy in the original decomposition of 24. Similarly, ${\cal L}_{(3)}$ involves just 8 terms, again a far cry from the total of 720. It is thus natural to assume that it would be possible to somehow identify and group the recurring expressions from the very beginning, in order to avoid any steps with great complexity.
\begin{figure}[h!]
     \includegraphics[width=0.2\textwidth]{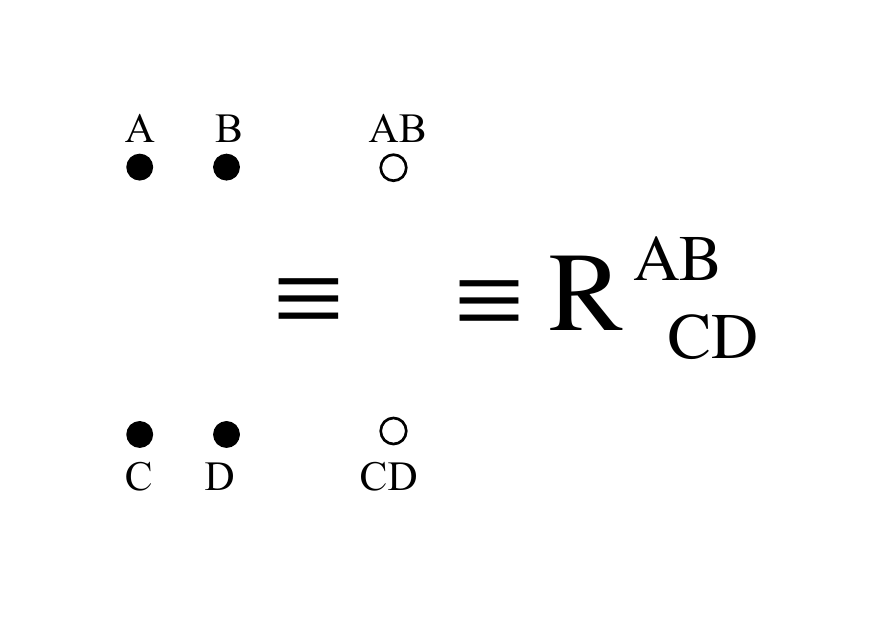} 
    \label{fig1}
\caption{Diagrammatic representation of the Riemann tensor in expanded (left) and compact (right) form.}
 \end{figure} 
Since we are primarily dealing with permutations of indices, it is natural to try to rephrase the problem in the language of permutation diagrams. These are often used in combinatorics to describe permutations in a graphical way. There is a number of variations in permutation diagrams, so we are going to adopt one of them which is more closely related to the problem in hand and helps to simplify our computations. In Figure 2, we have written down the possible permutation diagrams for the case of the Gauss-Bonnet term. ${\cal L}_{2}$ involves only expressions quadratic in the Riemann tensor. Since Lovelock densities are scalar quantities, all of the indices appearing in an expression have to be contracted, meaning that the same group of upper indices also appears as the corresponding set of down indices, only permuted.  To represent the possible combinations of index contractions, we write a Riemann tensor as a set of two upper and two lower dots, representing the respective upper and lower indices (see Fig. 1). 
  \begin{figure*}[h!]
\centering
     \includegraphics[width=0.8\textwidth]{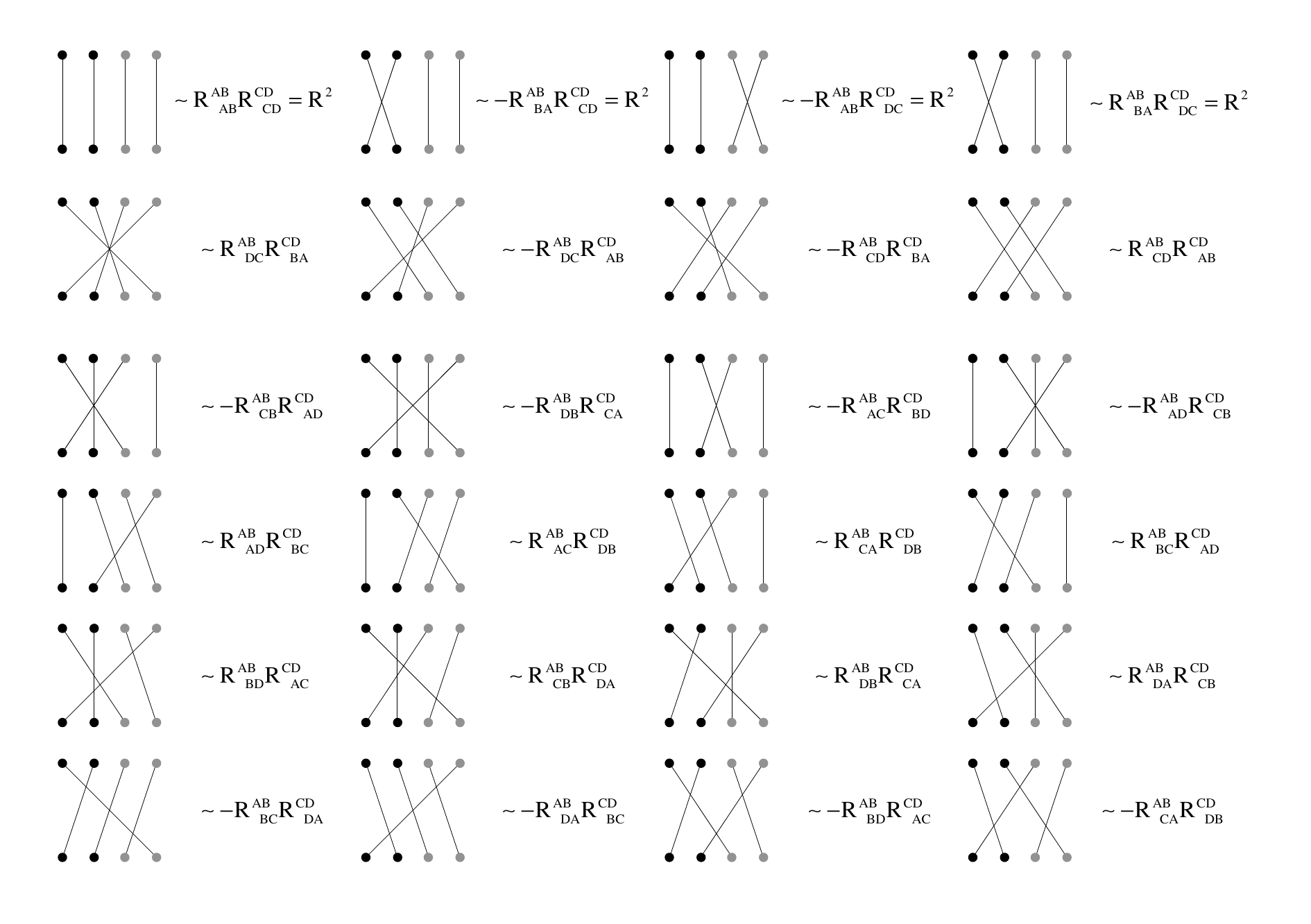} 
    \label{fig2}
\caption{Permutation diagrams for the Gauss-Bonnet scalar density. The first row corresponds to $R^{2}$ terms, the second to $R^{AB}_{\;\;CD}R^{CD}_{\;\;AB}$ and the final four are the $R^{A}_{\;B}R^{B}_{\;A}$ diagrams.}
 \end{figure*} 
A quadratic term is represented as two such sets side by side. For each contraction, we write a line connecting the corresponding dots of an upper and a lower index. In this way, writing two vertical lines between the dots of a single Riemann tensor yields the Ricci scalar, while two crossed lines gives $-R$. We can use these permutation graphs to identify the terms in component notation, using the following rules: 
\begin{enumerate}
\item Connect the dots corresponding to the indices of the Riemann tensors in all possible combinations, thus obtaining all contractions.
\item From each diagram, write the corresponding quadratic term. The upper indices are normally ordered, while the lower ones are determined by rearranging the upper indices according to the lines representing the permutation.
\item Write a minus sign if there is an odd number of line intersections in the permutation diagram. Do not count intersections with vertical lines, as these correspond to internal contractions and the corresponding indices cannot be exchanged any more.  
\end{enumerate}
These rules are applied to derive and transcribe all the diagrams giving the terms of the Gauss-Bonnet density in Fig. 2. Notice that the set of diagrams breaks down into three distinct groups, clearly visible by inspection in terms of symmetry. It is also immediately apparent that the $R^{A}_{\;B}R^{B}_{\;A}$ terms are four times more than those for $R^{2}$ and $R^{AB}_{\;\;CD}R^{CD}_{\;\;AB}$. From Fig. 2 we recover the well known expression ${\mathcal L}_{(2)}=(R^{ABCD} R_{ABCD}  - 4R^{AB} R_{AB}  +R^2)\theta ^*$, once we sum over all terms and take into account an overall $1/4$ factor in the definition of ${\mathcal L}_{(2)}$. We should also point out that the first diagram in the second row of Fig.2 in fact contains 2 overlapping intersections, since two pairs of lines intersect each other, hence bears no minus sign in agreement with the above rules. The two intersection points become visible once we draw the diagram in a non-symmetric way.    
\begin{figure}[b!]
\centering
     \includegraphics[width=0.5\textwidth]{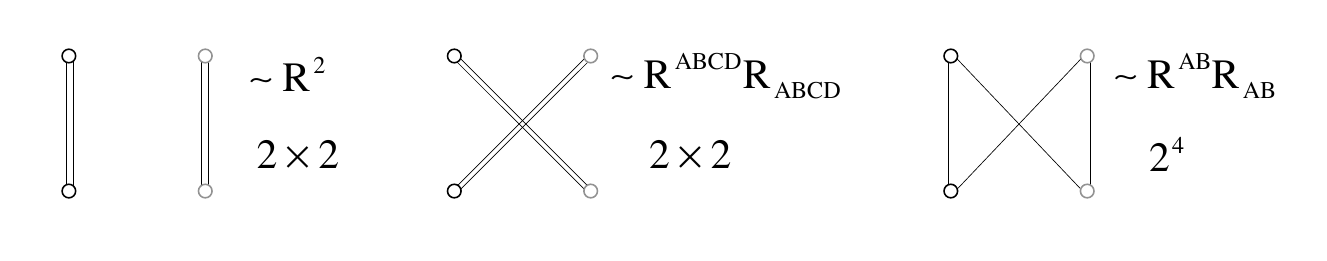} 
    \label{fig3}
\caption{The three distinct graphs giving rise to the corresponding terms of the Gauss-Bonnet density, with their respective combinatoric factors.}
 \end{figure} 
We see that the permutation diagram serves both as a book-keeping device to arrange the indices of the quadratic terms, as well as takes care of the minus signs coming with each odd permutation. The calculation of ${\mathcal L}_{(2)}$ is already considerably more transparent in this way and the emergence of the three different expressions in the final result becomes clear from symmetry considerations. Still, it doesn't cut down the paperwork needed. Just as before, we again had to write down all 24 terms, including many redundant ones. Fortunately, the fact that the pairs of upper and lower indices of the Riemann tensor are antisymmetric allows us to simplify our notation even further. It is easy to see that we can contract all indices of the same tensor by drawing two parallel or intersecting lines. The intersecting case corresponds to flipping the lower indices, so we get as a result $-R^{2}$ due to antisymmetry. But the minus sign is eliminated, since we also have another minus from the fact that this is an odd permutation, hence the result is the same as in the parallel lines diagram. We can thus account for both by just writing down the first diagram and multiply the result by two. This property is in fact more general and can be used to rewrite the Riemann tensor as a set of only one upper and one lower dot (Fig. 1). Each dot accounts for both indices and must be connected with two lines to be fully contracted. For each such vertex, there are two implicit possible ways in which the lines can be connected to the internal, hidden indices, depending on whether they intersect in the interior of the dot or not. Consequently, every such vertex should be multiplied by a factor of two, whenever the two lines are not both contracted with the corresponding dot of the same tensor. If a tensor only has internal contractions, we just multiply this expression by an overall 2, not 2 for each vertex. This is due to the symmetry of the Riemann under exchange of the two pairs of indices. 
\begin{figure*}[h!]
\centering
     \includegraphics[width=0.7\textwidth]{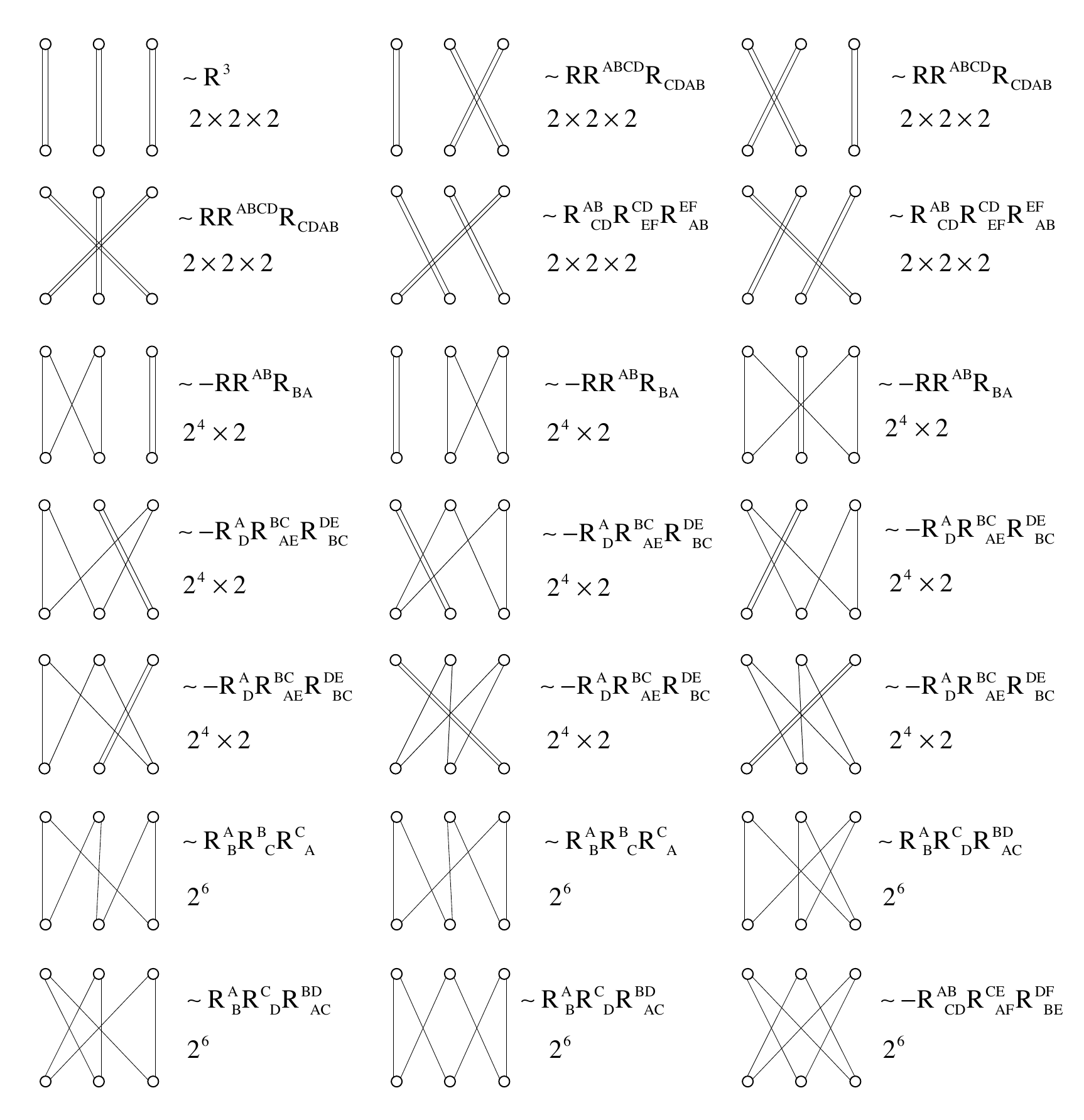} 
    \label{fig4}
\caption{Permutation diagrams (in compact notation) corresponding to the terms of the third order Lovelock density ${\mathcal L}_{(3)}$ and their combinatoric factors.}
 \end{figure*}
Using the strategy outlined above, we end up in the case of $L_{(2)}$ with only 3 distinct diagrams, matching the only terms appearing in the final expression of the Gauss-Bonnet scalar. The overall combinatoric factors and the minus sign of the $R^{AB} R_{AB}$ are all accounted for by the diagrams. Apart from a concise way to write down $L_{(2)}$, these compact permutation diagrams enable us to attack the considerably more difficult task of deriving $L_{(3)}$. Us it turns out, this can be done using only 21 compact diagrams in total (Fig. 4). Again, there is still an amount of degeneracy when considering the only 8 terms of the final result, still we do get a dramatic simplification compared to the original 720 terms one would have to deal with using the algebraic approach. Like terms can again be identified as originating from diagrams of particular symmetries and similar structure, which are easy to identify and group by inspection. The rules we use to construct the diagrams are very similar to the ones we discussed before for the Gauss-Bonnet term:
\begin{enumerate}
\item Connect all the dots between them, such that each has two lines attached to it.
\item From each diagram, write the corresponding third order term by rearranging the lower indices following the connections in the graph. Vertical lines correspond to internal contractions, one line will result in a Ricci tensor, two in the Ricci scalar.
\item Write a minus sign if there is an odd number of line intersections in the permutation diagram. Do not count intersections with vertical lines. 
\end{enumerate}
Combining these results and multiplying by a factor of $1/8$ we recover the expression ${\mathcal L}_{\left( 3 \right)}  = ( R^3  + 3RR^{ABCD} R_{CDAB}  - 12RR^{AB} R_{AB}  + 24R^A _{\;B} R^{BC} _{\;\;AD} R^D _{\;C}  + 16R^A _{\;B} R^B _{\;C} R^C _{\;A}  - 8R^{AB} _{\;\;CD} R^{CE} _{\;\;AF} R^{DF} _{\;\;BE}  + 2R^{AB} _{\;\;CD} R^{CD} _{\;\;EF} R^{EF} _{\;\;AB}  - 24R^A _{\;D} R^{BC} _{\;\;AE} R^{DE} _{\;\;BC} )\theta ^* $, in agreement with the algebraic result~\cite{MuellerHoissen:1985mm}. As we see, in the compact diagram notation the calculation becomes considerably faster and straightforward. 

One could proceed in a similar fashion to calculate even higher Lovelock densities in the compact formalism. The next term $L_{(4)}$ will involve contractions between four Riemann tensors. In total, it includes 25 distinct expressions. We immediately see that the first sixteen diagrams are exactly the same as the ones we encountered in the initial derivation of the Gauss-Bonnet scalar. The only thing we have to do is switch from single to double lines. Of course, just like for ${\mathcal L}_{(3)}$, we must also consider diagrams were lines do not originate and end on the same pair of dots, which become more involved. Still, the number of diagrams one has to deal with is certainly much lower than the total of $8!$ terms in the full expansion of ${\mathcal L}_{(4)}$. It is also reasonable to assume that further reduction of this redundancy can be achieved by identifying particular patterns between the form of the diagrams and the terms they give rise to.   

\section*{Acknowledgments}
The author would like to thank Christos Charmousis for stimulating discussions and for taking the time to read the manuscript and provide feedback. This work is supported by the CNRS and the Universit\'e de Paris-Sud XI. 
   
\bibliographystyle{unsrt}
\bibliography{references}

\end{document}